\documentclass[cameraready]{Interspeech}
\title{\texorpdfstring{Latency-Configurable Streaming Speech Enhancement\\via Asymmetric Temporal Padding}{Latency-Configurable Streaming Speech Enhancement via Asymmetric Temporal Padding}}

\author[affiliation={1}]{Yunsik}{Kim}
\author[affiliation={1,2}, correspondingauthor]{Yoonyoung}{Chung}

\address{
    $^1$ Department of Electrical Engineering, Pohang University of Science and Technology (POSTECH), Pohang 37673, Republic of Korea \\
    $^2$ Intus Co. Ltd., Pohang 37673, Republic of Korea
}

\email{yskim@postech.ac.kr; ychung@postech.ac.kr}


\usepackage{amsmath}
\usepackage{amssymb}
\usepackage{multirow}
\usepackage{booktabs}
\usepackage{graphicx}

\begin{document}

\maketitle

\begin{abstract}
    Streaming speech enhancement requires balancing algorithmic latency against quality, yet existing approaches largely treat this as a binary causal versus non-causal choice. LaCo-SENet addresses this issue with two mechanisms parameterized by a single training-time hyperparameter. First, asymmetric temporal padding redistributes past and future context in convolutions, enabling systematic latency configuration. Second, dual-buffer streaming combines state buffers for past context with lookahead buffers that supply future context at both the input and feature levels. Selective state updates also prevent future-frame leakage into the streaming state, ensuring training--inference consistency. On VoiceBank+DEMAND, a fixed-budget (1.37M parameters) backbone yields a family of models spanning 12.5--75.0 ms, with PESQ rising from 3.35 to 3.43. At just 12.5 ms (fully causal), a PESQ of 3.35 matches or exceeds the prior causal state-of-the-art (3.27 at 46.5 ms).
\end{abstract}

\section{Introduction}
\label{sec:intro}

Streaming speech enhancement is essential for real-time applications such as telephony, conferencing, hearing aids, and on-device voice interfaces, where algorithmic latency directly governs system responsiveness~\cite{reddy2021icassp_dns, schroter2022lowlatency_ha}.
Within the 10--80\,ms regime where most applications operate, quality generally improves with additional lookahead, yet existing streaming models each target a fixed latency point. No prior work has proposed a unified framework for exploring this trade-off within a single convolutional architecture.
Redistributing convolution padding asymmetrically between past and future context provides a practical latency-configuration knob that preserves the receptive field and parameter count.
However, deploying per-layer asymmetric padding in chunk-based streaming introduces a state corruption problem. Lookahead frames recorded into convolution state buffers are replayed in subsequent chunks, progressively distorting outputs.
Our dual-buffer streaming framework resolves this through selective state updates that restrict state recording to current-chunk frames.

We present \textbf{LaCo-SENet} (Latency-Configurable SE Network, 1.37M parameters, built on PrimeK-Net~\cite{primeknet2025}). At just \textbf{12.5\,ms} (fully causal), it achieves PESQ \textbf{3.35} on VoiceBank+DEMAND, matching or exceeding the best reported causal PESQ of 3.27 at 46.5\,ms~\cite{atennuate2025pei}. Varying a single training-time hyperparameter trades latency for quality, reaching PESQ 3.43 at 75.0\,ms.

\begin{figure*}[t]
\centering
\includegraphics[width=0.85\textwidth]{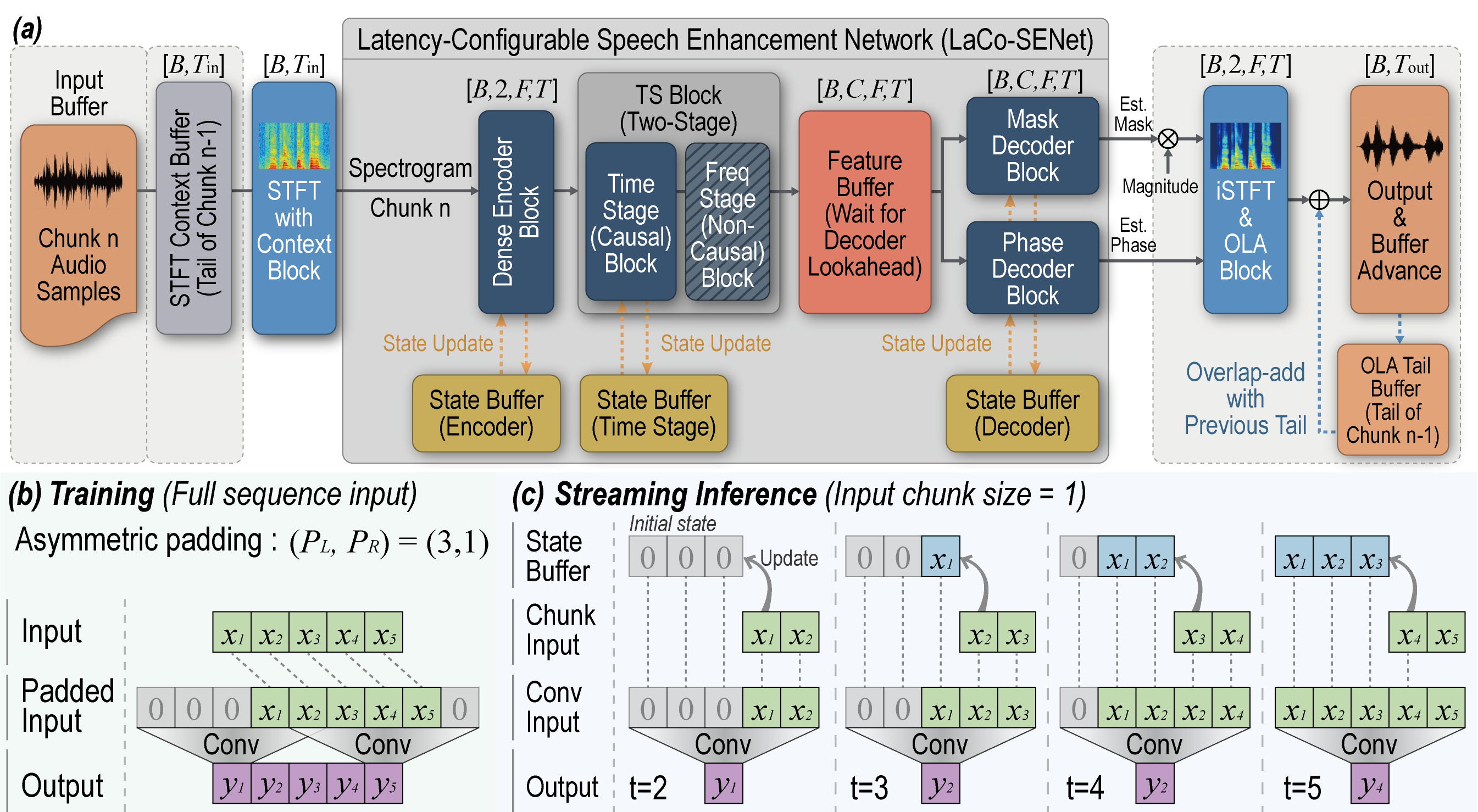}
\caption{Overview of LaCo-SENet. (a) Dual-buffer streaming architecture: the STFT context buffer provides encoder lookahead at the input level; the feature buffer provides decoder lookahead; and state buffers preserve past context. (b) Training with asymmetric temporal padding $(P_L, P_R)$, redistributing past and future context while preserving the receptive field. (c) Streaming inference with chunk-wise processing (chunk size = 1): selective state updates prevent lookahead frames from corrupting convolution states.}
\label{fig:overview}
\end{figure*}

Our contributions are threefold: (1) we show that \textbf{asymmetric temporal padding} serves as a practical, training-time latency-configuration knob for streaming convolutional SE, enabling systematic exploration of discrete latency--quality trade-offs without changing the receptive field, parameter count, or architecture; (2) we propose a \textbf{dual-buffer streaming framework} with selective state updates that preserves equivalence between full-sequence and chunk-wise inference by preventing future-frame leakage into the streaming state; and (3) we show that a fixed 1.37M-parameter architecture \textbf{spans 12.5--75.0\,ms latency} (PESQ 3.35--3.43) by training with different padding ratios, matching or exceeding prior causal models at lower latency.

\section{Related Work}
\label{sec:related}

\noindent\textbf{Streaming SE architectures.}
Diverse approaches address streaming speech enhancement, including DSP-DNN hybrids~\cite{rnnoise2018valin}, complex-valued convolutional-recurrent networks~\cite{dccrn2020hu}, perceptual deep filtering~\cite{deepfilternet3_2023schroter}, raw-waveform state-space autoencoders~\cite{atennuate2025pei}, Mamba-based models~\cite{semamba2024chao}, and xLSTM-based designs~\cite{xlstmsenet2025kuhne}.
Each is locked to a single latency point by its design, and none can systematically trade latency for quality within a single architecture and parameter budget.

\noindent\textbf{Stateful convolution and lookahead.}
Maintaining temporal context across chunk boundaries is fundamental to streaming.
FastWaveNet~\cite{fastwavenet2016lepaine} and cached convolution~\cite{cachedconv2022caillon} cache past activations to replace zero left-padding, while Denoiser~\cite{denoiser2020defossez} applies a similar mechanism in waveform-domain models.
For future context, Emformer~\cite{emformer2021shi} and Stateful Conformer~\cite{statefulconformer2024noroozi} add right-context to attention-based architectures through segment-level masking or memory policies.
These approaches address past and future context individually but do not consider their interaction under per-layer asymmetric padding.

\noindent\textbf{Per-layer asymmetric padding.}
Among convolutional streaming approaches, existing methods constrain where future context may enter.
DeepFilterNet~\cite{deepfilternet3_2023schroter} supplies lookahead through a global input shift while keeping every convolution purely causal; cached convolution~\cite{cachedconv2022caillon} converts symmetric padding into an equivalent fixed output delay.
Both constraints deliberately preclude per-layer asymmetric padding, avoiding state corruption at the cost of latency flexibility. Their lookahead is therefore either fixed at the input boundary or set entirely by the symmetric kernel size.
Our approach distributes padding asymmetrically within each layer, enabling discrete latency configuration while resolving the resulting state corruption through selective updates.

\section{Proposed Method}
\label{sec:method}

\subsection{Asymmetric temporal padding}
\label{sec:asympad}

We fix the total temporal padding while distributing it asymmetrically between past and future. Let $P$ denote the per-side padding of a standard symmetric convolution, so that the total padding is $P_{\text{tot}} = 2P$.
Defining $\texttt{padding\_ratio}$ as $\mathbf{r}=(r_L,r_R)$ with $r_L+r_R=1$, the temporal padding is distributed as:
\begin{equation}
P_L = \mathrm{round}(P_{\text{tot}} \cdot r_L),\quad
P_R = P_{\text{tot}} - P_L.
\end{equation}
For a given operating point, $\mathbf{r}$ is a fixed training-time hyperparameter shared by all asymmetric convolution layers; it is not adapted to SNR, input content, or runtime conditions.

Frequency-axis padding remains symmetric, and only the time axis receives the asymmetric split $(P_L,P_R)$. Since $P_{\text{tot}}$ is fixed, the receptive field size is preserved. Only the ratio of past-to-future access changes, enabling quantized latency configuration. Although the padding redistribution is algebraically straightforward, deploying it in a stateful streaming pipeline is non-trivial. Every layer with $P_R{>}0$ introduces future frames into the convolution window, and naively caching all frames corrupts subsequent chunks (Section~\ref{sec:lacosenet}).

\begin{table*}[!t]
\centering
\small
\setlength{\abovecaptionskip}{2pt}
\setlength{\belowcaptionskip}{0pt}
\renewcommand{\arraystretch}{0.88}
\setlength{\aboverulesep}{0.2ex}
\setlength{\belowrulesep}{0.2ex}
\caption{Comparison on VoiceBank+DEMAND. Best causal result per metric in \textbf{bold}; models marked $\dagger$ or $\ddagger$ excluded from bold ranking. ---: not reported or unverifiable.}
\label{tab:results}
\begin{tabular}{@{}l c r r r r r r r@{}}
\toprule
Model & Year & $\tau$ (ms) & Params & PESQ & STOI & CSIG & CBAK & COVL \\
\midrule
Noisy & --- & --- & --- & 1.97 & .921 & 3.35 & 2.44 & 2.63 \\
\midrule
RNNoise~\cite{rnnoise2018valin}        & 2018 & 10            & 0.06M & 2.33 & .922 & 3.40 & 2.51 & 2.84 \\
GaGNet~\cite{gagnet2022li}             & 2022 & ${\sim}$10    & 5.94M & 2.94 & ---  & 4.26 & 3.45 & 3.59 \\
DFNet3~\cite{deepfilternet3_2023schroter} & 2023 & 40         & 2.13M & 3.17 & .944 & 4.34 & 3.61 & 3.77 \\
aTENNuate~\cite{atennuate2025pei}      & 2025 & 46.5          & 0.84M & 3.27 & ---  & 4.57 & 2.85 & 3.96 \\
xLSTM-SENet$^\dagger$~\cite{xlstmsenet2025kuhne} & 2025 & ---  & ${\sim}$2.2M & 3.26 & .950 & 4.57 & 3.79 & 4.00 \\
SEMamba$^\dagger$~\cite{semamba2024chao} & 2024 & ---           & 1.41M & 3.29 & .950 & ---  & ---  & ---  \\
\midrule
LaCo-SENet\,{\scriptsize($L_{\text{enc}}{=}0,\,L_{\text{dec}}{=}0$)}   & \multirow{5}{*}{2026} &  12.5  & \multirow{5}{*}{1.37M} & 3.35{\scriptsize$\pm$.02} & .952{\scriptsize$\pm$.000} & 4.61{\scriptsize$\pm$.01} & 3.71{\scriptsize$\pm$.01} & 4.05{\scriptsize$\pm$.02} \\
LaCo-SENet\,{\scriptsize($L_{\text{enc}}{=}1,\,L_{\text{dec}}{=}1$)}   &  &  25.0  &  & 3.36{\scriptsize$\pm$.01} & .953{\scriptsize$\pm$.000} & 4.62{\scriptsize$\pm$.01} & 3.72{\scriptsize$\pm$.02} & 4.07{\scriptsize$\pm$.01} \\
LaCo-SENet\,{\scriptsize($L_{\text{enc}}{=}3,\,L_{\text{dec}}{=}3$)}   &  &  50.0  &  & 3.40{\scriptsize$\pm$.02} & .953{\scriptsize$\pm$.001} & 4.63{\scriptsize$\pm$.02} & 3.72{\scriptsize$\pm$.01} & 4.09{\scriptsize$\pm$.02} \\
LaCo-SENet\,{\scriptsize($L_{\text{enc}}{=}5,\,L_{\text{dec}}{=}5$)}   &  &  75.0  &  & \textbf{3.43}{\scriptsize$\pm$.01} & \textbf{.954}{\scriptsize$\pm$.001} & \textbf{4.66}{\scriptsize$\pm$.02} & \textbf{3.78}{\scriptsize$\pm$.00} & \textbf{4.12}{\scriptsize$\pm$.02} \\
LaCo-SENet$^\ddagger$\,{\scriptsize($L_{\text{enc}}{=}15,\,L_{\text{dec}}{=}15$)} &  & 200.0  &  & 3.47{\scriptsize$\pm$.02} & .957{\scriptsize$\pm$.001} & 4.69{\scriptsize$\pm$.01} & 3.79{\scriptsize$\pm$.03} & 4.17{\scriptsize$\pm$.02} \\
\midrule
PrimeK-Net~\cite{primeknet2025}        & 2025 & ---           & 1.41M & 3.61 & ---  & 4.81 & 3.98 & 4.35 \\
\bottomrule
\end{tabular}
\vspace{1pt}
\parbox{\textwidth}{\raggedright\scriptsize
$^\dagger$\,Labeled unidirectional, but encoder convolutions use non-causal symmetric padding; algorithmic latency not deterministically computable.\\
$^\ddagger$\,Symmetric padding ($r_R{=}0.5$); architecture upper bound, excluded from best-causal ranking.}
\end{table*}

\subsection{Dual-buffer streaming framework}
\label{sec:lacosenet}

In the streaming setting, the input spectrogram is processed in chunks of $C$ time frames. Let $k$ denote the chunk index and $\ell$ a convolution layer with asymmetric temporal padding $(P_L^{(\ell)}, P_R^{(\ell)})$. The total right-side padding accumulated across the encoder and decoder determines the algorithmic latency:
\begin{equation}
\begin{aligned}
L_{\text{enc}} &= \sum_{\ell\in \mathcal{E}} P_R^{(\ell)},\\
L_{\text{dec}} &= \max_{b\in\mathcal{B}}
\sum_{\ell\in \mathcal{D}_b} P_R^{(\ell)},\quad
\mathcal{B}=\{\mathrm{mask},\mathrm{phase}\},
\end{aligned}
\end{equation}
\begin{equation}
\tau_{\mathrm{ms}} = 1000 \cdot \frac{(L_{\text{enc}} + L_{\text{dec}}) \cdot h + W/2}{f_s},
\end{equation}
where $\mathcal{E}$ denotes the asymmetric convolution layers in the encoder, $\mathcal{D}_b$ those in decoder branch $b$, $h$ is the hop size, $W$ the STFT window size, and $f_s$ the sample rate. The term $W/2$ accounts for the STFT center delay. Since the mask and phase decoders run in parallel, decoder lookahead is maximized across branches rather than summed across them.
Deploying asymmetric padding in chunk-based streaming raises three challenges, which we address below.

\noindent\textbf{Challenge 1: State buffer for past context.}
At chunk boundaries, convolution left-padding is filled with zeros rather than actual past activations, which diverges from full-sequence processing~\cite{cachedconv2022caillon, fastwavenet2016lepaine}.
To resolve this, each layer $\ell$ with left padding $P_L^{(\ell)}$ caches the last $P_L^{(\ell)}$ frames of the current input as state $\mathbf{s}^{(\ell)}_{k}$. At chunk $k$, zero left-padding is replaced by this cache:
\begin{equation}
\begin{aligned}
\tilde{\mathbf{x}}^{(\ell)}_k &= \big[\, \mathbf{s}^{(\ell)}_{k-1};\ \mathbf{x}^{(\ell)}_k;\ \mathbf{0}\,\big],\\
\mathbf{s}^{(\ell)}_{k} &\leftarrow \text{last } P_L^{(\ell)} \text{ frames of } \mathbf{x}^{(\ell)}_k.
\end{aligned}
\end{equation}

\noindent\textbf{Challenge 2: Lookahead buffer for future context.}
When $P_R{>}0$, convolutions reference frames beyond the chunk boundary. In streaming, these are unavailable unless explicitly buffered~\cite{emformer2021shi}.
LaCo-SENet delays its output and buffers additional input at two levels.

\noindent\textit{Input lookahead (encoder).}
When $L_{\text{enc}}>0$, the encoder input is extended to $C+L_{\text{enc}}$ frames, so that every encoder-side convolution accesses actual future frames instead of zeros.

\noindent\textit{Feature buffer (decoder).}
When $L_{\text{dec}}>0$, encoder output features are accumulated in a buffer, and the decoder is invoked only once the buffer satisfies $\text{buffered\_frames}\ge C + L_{\text{dec}}$.
Upon invocation, the decoder receives an extended feature sequence. Only the portion corresponding to the current chunk is emitted, while the remaining frames serve as future context. Since at least $L_{\text{dec}}$ lookahead frames lie between the emitted region and the zero right-padding boundary, all emitted outputs are computed with actual future context rather than zeros.

\noindent\textbf{Challenge 3: Selective state update.}
When both state and lookahead buffers are active, a subtler problem emerges. If the state buffer records all frames, including the appended lookahead, those frames appear twice in subsequent chunks (once from the state and once as newly arrived input), progressively distorting outputs.
LaCo-SENet restricts the state-update scope via a selection operator $\Pi_C(\mathbf{x}) = \mathbf{x}_{1:C}$ that keeps only current-chunk frames.
Even when processing an extended input $\mathbf{x}_{k,\text{ext}}$ that contains $C$ current frames plus lookahead, the state update depends only on the current-chunk frames:
\begin{equation}
\mathbf{s}^{(\ell)}_{k} \leftarrow \text{last } P_L^{(\ell)} \text{ frames of } \Pi_C(\mathbf{x}^{(\ell)}_{k,\text{ext}}).
\end{equation}
Consequently, lookahead frames participate in the forward computation but are never recorded into the state.

\subsection{Backbone architecture}
\label{sec:architecture}

Let $x[t]\in\mathbb{R}$ denote a single-channel noisy waveform. We compute the STFT $X_{f,n}\in\mathbb{C}$, apply power-law magnitude compression ($c{=}0.3$), and represent the input as $\mathbf{X}\in\mathbb{R}^{2\times T\times F}$ (compressed magnitude and phase).

Our backbone $g_\theta$ follows PrimeK-Net~\cite{primeknet2025} with streaming-specific modifications. We decompose it into an encoder $E_\theta$, a sequence of time--frequency blocks $\mathcal{T}_\theta$, and two parallel decoders $D^M_\theta$ (mask) and $D^\Phi_\theta$ (phase):
\begin{equation}
\mathbf{H} = \mathcal{T}_\theta(E_\theta(\mathbf{X})),\quad
\hat{\mathbf{M}} = D^M_\theta(\mathbf{H}),\quad
\hat{\mathbf{\Phi}} = D^\Phi_\theta(\mathbf{H}).
\end{equation}
The backbone predicts a magnitude mask $\hat{\mathbf{M}}\in\mathbb{R}^{T\times F}$ and an enhanced phase $\hat{\mathbf{\Phi}}\in\mathbb{R}^{T\times F}$, from which the enhanced spectrogram is reconstructed via masking and phase replacement.
The input is projected to $C_h$ channels, processed by the encoder's Dense Dilated Depthwise Block (DSDDB), and downsampled along frequency (stride $(1,2)$), yielding $\mathbf{H}\in\mathbb{R}^{C_h\times T\times F'}$ ($F'{=}\lceil F/2 \rceil$). Each decoder applies a DSDDB followed by transposed convolution to restore frequency resolution.

\subsubsection{Dense Dilated Depthwise Block (DSDDB)}

DSDDB stacks four depthwise separable convolutions with exponentially increasing dilation rates $(1,2,4,8)$ and dense connections. Each subsequent layer takes the concatenation of all preceding layer outputs as input. The depthwise convolution uses AsymmetricConv2d (Section~\ref{sec:asympad}) so that the temporal padding ratio $\mathbf{r}$ consistently controls the allocation of past and future context across the network.

\subsubsection{Time--Frequency Sequence Block (TS Block)}

Each TS Block processes time and frequency axes sequentially. For batch size $N$, the time branch reshapes the feature map from $\mathbb{R}^{N\times C_h\times T\times F'}$ to $\mathbb{R}^{(NF')\times C_h\times T}$ and applies a stack of Channel Attention Blocks (CAB) and Group Prime Kernel FFN (GPKFFN) modules, all employing causal convolutions. The frequency branch reshapes it to $\mathbb{R}^{(NT)\times C_h\times F'}$ and applies the same block types non-causally, since the frequency axis carries no temporal ordering.

\subsubsection{Streaming-specific modifications}

The following changes are required to make the backbone compatible with chunk-based streaming:
\begin{enumerate}
\item \textbf{Normalization}: InstanceNorm $\to$ BatchNorm. InstanceNorm statistics depend on the full sequence length, making them inconsistent across chunk sizes. BatchNorm uses fixed running statistics and is chunk-size invariant.
\item \textbf{Channel attention (SCA)}: AdaptiveAvgPool1d (global) $\to$ CausalConv1d (depthwise, $K_{\text{sca}}{=}\texttt{sca\_kernel\_size}$). Global pooling aggregates future frames, violating causality. A causal depthwise convolution restricts the context to past frames only.
\end{enumerate}

\subsubsection{Model configuration}

The backbone uses dense channels $C_h{=}64$, DSDDB depth 4, four TS Blocks (2 time + 2 freq), time kernels $[3,5,7,11]$, frequency kernels $[3,11,23,31]$, SCA kernel size 11, and STFT parameters (win/hop/fft) of 400/100/400 samples (25.0/6.25/25.0\,ms at 16\,kHz).

\section{Experiments}
\label{sec:experiments}

\subsection{Experimental Setup}
\label{sec:setup}

We evaluated on VoiceBank+DEMAND~\cite{valentini2017noisy} at 16\,kHz, comprising 11,572 training and 824 test utterances mixed with 10 noise types at four SNR levels.

\noindent\textbf{Training.}
We used AdamW ($\text{lr}{=}5{\times}10^{-4}$, $(\beta_1,\beta_2){=}(0.8,0.99)$) with exponential LR decay, batch size~8, and 400K~steps.
The training loss follows PrimeK-Net~\cite{primeknet2025}: $\mathcal{L} = 0.9\mathcal{L}_{\mathrm{mag}} + 0.3\mathcal{L}_{\mathrm{pha}} + 0.1\mathcal{L}_{\mathrm{com}} + 0.05\mathcal{L}_{\mathrm{con}} + 0.05\mathcal{L}_{\mathrm{gan}}$, weighting magnitude, phase, complex, consistency, and MetricGAN~\cite{metricgan2019fu} metric terms.
Each latency configuration was trained independently with three random seeds. The best checkpoint was selected by validation PESQ, and we report mean $\pm$ s.d.\ across seeds.

\noindent\textbf{Evaluation.}
We report PESQ (wideband), STOI, and the composite measures CSIG, CBAK, COVL~\cite{hu2007evaluation} on full-length test utterances using full-sequence inference. Streaming equivalence is verified in Section~\ref{sec:ablation}.

\noindent\textbf{Latency configurations.}
Encoder and decoder share a common padding ratio $\mathbf{r}{=}(r_L,r_R)$, giving equal lookahead ($L_{\text{enc}}{=}L_{\text{dec}}$). Sweeping $r_R$ from 0 (fully causal) to 0.5 (symmetric) yields nine configurations spanning $\tau{=}12.5$--$200.0$\,ms. These comprise six at $L_{\text{enc}}{=}L_{\text{dec}}\in\{0,\dots,5\}$, two intermediate (100.0, 150.0\,ms), and a symmetric reference ($r_R{=}0.5$) as an architecture upper bound. Table~\ref{tab:results} reports four representative causal points and the symmetric reference. Figure~\ref{fig:latency_pesq} shows all nine.

\noindent\textbf{Comparison models.}
Our comparison includes four causal baselines with verifiable latency spanning 10--46.5\,ms: RNNoise~\cite{rnnoise2018valin}, GaGNet~\cite{gagnet2022li}, DeepFilterNet3~\cite{deepfilternet3_2023schroter}, and aTENNuate~\cite{atennuate2025pei}, along with the non-causal PrimeK-Net~\cite{primeknet2025} as an upper bound.
Latency values are from original papers or estimated as $(\text{window}{-}\text{hop})/f_s$.
SEMamba~\cite{semamba2024chao} and xLSTM-SENet~\cite{xlstmsenet2025kuhne} report unidirectional variants but use non-causal symmetric encoder padding, so their algorithmic latency is indeterminate. They are excluded from latency-based ranking.

\subsection{Results}
\label{sec:results}

\begin{figure}[t]
\centering
\includegraphics[width=0.80\columnwidth]{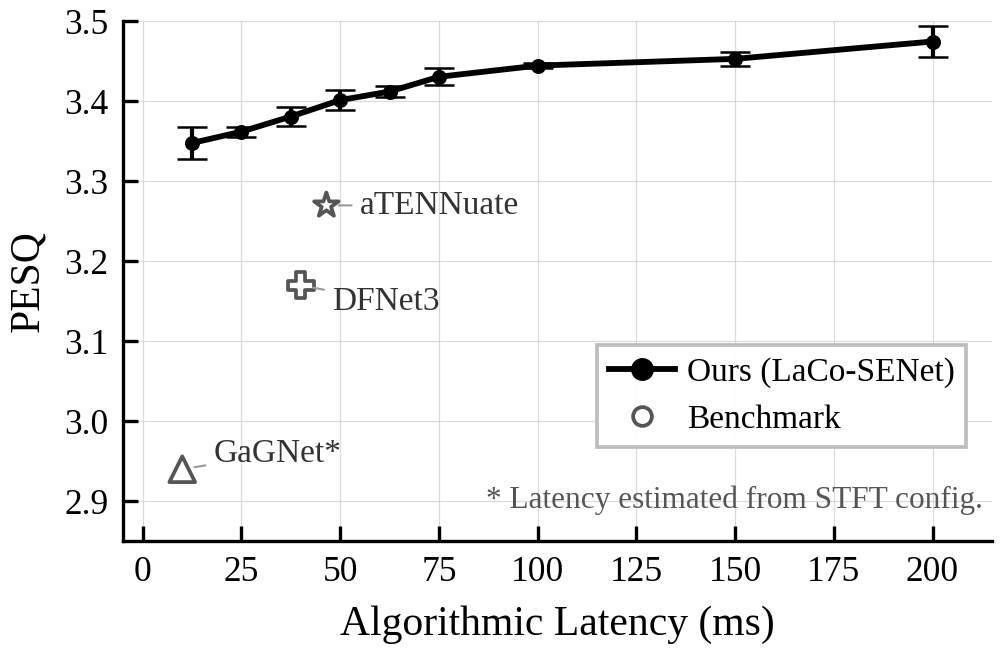}
\caption{PESQ vs.\ algorithmic latency on VoiceBank+DEMAND. LaCo-SENet (filled circles, connected) spans 12.5--200.0\,ms with a constant 1.37M parameters. Open markers denote selected comparison models shown in the plot.}
\label{fig:latency_pesq}
\end{figure}

At 12.5\,ms (fully causal), LaCo-SENet achieves PESQ 3.35, outperforming all causal baselines including aTENNuate~\cite{atennuate2025pei} (3.27 at 46.5\,ms).
PESQ then rises with lookahead (Figure~\ref{fig:latency_pesq}), from 3.35 at 12.5\,ms to 3.43 at 75.0\,ms, with the largest gains in the 25--75\,ms range and diminishing returns beyond. The symmetric reference adds only $+$0.04 over a further 125\,ms, reaching 3.47 at 200.0\,ms. The streaming configurations thus retain 93--95\% of the non-causal PrimeK-Net~\cite{primeknet2025} quality (3.61) at comparable parameter count (1.37M vs.\ 1.41M).

\begin{figure}[t]
\centering
\includegraphics[width=0.80\columnwidth]{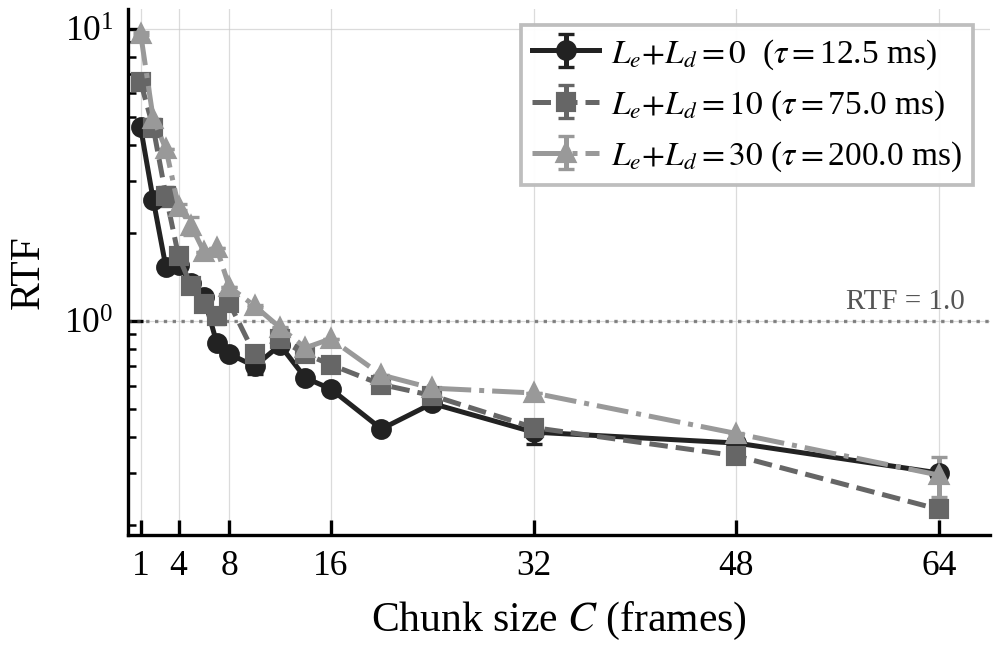}
\caption{Steady-state streaming RTF vs.\ chunk size $C$ (frames) for three total lookahead values $L_{\text{enc}}{+}L_{\text{dec}} \in \{0,10,30\}$.}
\label{fig:rtf_chunksize}
\end{figure}

\noindent\textbf{Streaming throughput.}
We measured end-to-end \emph{steady-state} streaming RTF (STFT + model + iSTFT) using ONNX Runtime on a single thread of an Intel Xeon Silver 4510.
Increasing $C$ amortizes fixed per-step overhead (Figure~\ref{fig:rtf_chunksize}). RTF falls from 4.59 at $C{=}1$ to 0.30 at $C{=}64$ for $L_{\text{enc}}{+}L_{\text{dec}}{=}0$, while the large-lookahead penalty narrows from 2.10$\times$ at $C{=}1$ to ${\approx}1.0\times$ by $C{=}64$ (RTF ${\approx}$0.23--0.30 across all configurations).
Real-time operation (RTF${<}1$) needs $C{\ge}7$--$12$. At $C{=}8$ the total latency (algorithmic plus chunk buffering) is 62.5\,ms for the causal case, at RTF 0.77.

\begin{table}[!t]
\centering
\small
\setlength{\abovecaptionskip}{2pt}
\setlength{\belowcaptionskip}{0pt}
\setlength{\aboverulesep}{0.2ex}
\setlength{\belowrulesep}{0.2ex}
\caption{Ablation on selective state update (SSU). Streaming PESQ at chunk size $C{=}1$ with SSU enabled vs.\ disabled; tail 30 frames trimmed to exclude end-of-utterance OLA boundary artifacts. Mean $\pm$ std over three seeds.}
\label{tab:ablation}
\begin{tabular}{@{}c c c c c@{}}
\toprule
$L_{\text{enc}}{+}L_{\text{dec}}$ & $\tau$ (ms) & w/ SSU & w/o SSU & $\Delta$PESQ \\
\midrule
0  & 12.5  & 3.39{\scriptsize$\pm$.01} & 3.39{\scriptsize$\pm$.01} & 0.00 \\
2  & 25.0  & 3.41{\scriptsize$\pm$.01} & 1.86{\scriptsize$\pm$.31} & $-$1.56 \\
6  & 50.0  & 3.45{\scriptsize$\pm$.01} & 1.45{\scriptsize$\pm$.14} & $-$2.00 \\
10 & 75.0  & 3.48{\scriptsize$\pm$.01} & 1.39{\scriptsize$\pm$.01} & $-$2.09 \\
30 & 200.0 & 3.52{\scriptsize$\pm$.02} & 2.04{\scriptsize$\pm$.06} & $-$1.48 \\
\bottomrule
\end{tabular}
\end{table}

\subsection{Ablation Study}
\label{sec:ablation}

Table~\ref{tab:ablation} evaluates SSU by disabling it during streaming inference ($C{=}1$), with $L_{\text{tot}}{=}L_{\text{enc}}{+}L_{\text{dec}}$.
For all asymmetric configurations ($L_{\text{tot}}{=}2$--$10$), disabling SSU drops PESQ below the noisy baseline (1.97), worsening monotonically with lookahead, from $-1.56$ at $L_{\text{tot}}{=}2$ to $-2.09$ at $L_{\text{tot}}{=}10$. The symmetric reference ($L_{\text{tot}}{=}30$, $r_R{=}0.5$) drops less ($-1.48$), as its equal left--right padding halves the state buffer, reducing corruption. SSU is thus essential whenever asymmetric padding introduces lookahead. We further verify that chunk-wise streaming outputs are numerically identical to full-sequence outputs across all configurations.

\section{Conclusion}
\label{sec:conclusion}

We presented LaCo-SENet, a dual-buffer streaming framework whose algorithmic latency is configurable via asymmetric temporal padding, with training--inference equivalence preserved by selective state updates. On VoiceBank+DEMAND, a fixed 1.37M-parameter architecture spans 12.5--75.0\,ms (PESQ 3.35--3.43) across padding ratios, surpassing prior causal models at lower latency.

\section{Code Availability}
\label{sec:code}

The implementation is available at \url{https://github.com/yskim3271/LaCo-SENet}.

\section{Acknowledgments}
This work was supported by the National Research Foundation of Korea (NRF) grant funded by the Ministry of Science and ICT (RS-2025-00516311); by the Institute of Information \& Communications Technology Planning \& Evaluation (IITP) grant funded by the Korean government (RS-2019-II191906, Artificial Intelligence Graduate School Program); by the Regional Innovation System \& Education project (Specialized Industry Scale-UP unit), supported by Gyeongsangbuk-do; and by the High-Performance Computing Support Project funded by the Government of the Republic of Korea (Ministry of Science and ICT).

\section{Generative AI Use Disclosure}
Generative AI tools were used solely to edit and polish the manuscript text (e.g., grammar and wording). They were not used to generate research content---including the method, experiments, analysis, or results---and are not credited as authors.

\bibliographystyle{IEEEtran}
\bibliography{mybib}

\end{document}